\documentclass[apj]{emulateapj}
\usepackage{apjfonts}
\usepackage{graphicx}
\usepackage{amsfonts}
\usepackage{amsmath}
\usepackage{color}
\usepackage{amsbsy}
\usepackage{mathpazo,xfrac}
\usepackage{multirow}
\usepackage{wasysym}
\usepackage{url}
\usepackage{hyperref}

\newlength{\hfwidth}
\newlength{\hfwidthsingle}
\newcommand{\ptderiv}[1]{\frac{\partial{#1}}{\partial{t}}}

\newcommand{\ttimes}[1]{10^{#1}}

\newcommand{\vt}[1]{\boldsymbol{\mathrm{#1}}}       
\renewcommand{\v}[1]{{\boldsymbol{#1}}} 

\newcommand{\ksi}{\xi}

\newcommand{\del}{\v{\nabla}}
\newcommand{\grad}{\del}
\newcommand{\Div}{\del\cdot}

\newcommand{\cv}{c_v}
\newcommand{\cp}{c_p}

\newcommand{\Eq}[1]{Eq.~(\ref{#1})}

\newcommand{\Eqss}[2]{Eqs.~(\ref{#1})--(\ref{#2})}
\newcommand{\eq}[1]{\Eq{#1}}
\newcommand{\eqp}[1]{(Eq.~\ref{#1})}

\newcommand{\eqss}[2]{\Eqss{#1}{#2}}

\newcommand{\Fig}[1]{Fig.~\ref{#1}}
\newcommand{\fig}[1]{\Fig{#1}}

\definecolor{brown}{rgb}{0.42,0.24,0.07}
\definecolor{darkgreen}{rgb}{0.0,0.6,0.00}
\definecolor{purple}{rgb}{0.7,0.0,0.7}
\definecolor{black}{rgb}{0.0,0.0,0.0}

\def\white#1{\textcolor{white}{#1}}

\shorttitle{Convective overstability}
\shortauthors{Lyra}

\slugcomment{Draft version}

\begin{document}

\title{Convective overstability in accretion disks\\3D linear analysis and nonlinear saturation}

\author{Wladimir Lyra\altaffilmark{1,2,3}}
\email{wlyra@caltech.edu}
\altaffiltext{1}{Jet Propulsion Laboratory, California Institute of Technology, 4800 Oak Grove Drive, Pasadena, CA, 91109, USA}
\altaffiltext{2}{Department of Geology and Planetary Sciences, California Institute of Technology, 1200 E. California Ave, Pasadena CA, 91125}
\altaffiltext{3}{Sagan Fellow}

\begin{abstract}
Recently, Klahr \& Hubbard (2014) claimed that a hydrodynamical linear overstability
exists in protoplanetary disks, powered by buoyancy in the presence of thermal
relaxation. We analyse this claim, confirming it through rigorous
compressible linear analysis. We model the system numerically,
reproducing the linear growth rate for all cases studied. We also
study the saturated properties of the overstability in the shearing box, finding that the saturated state
produces finite amplitude fluctuations strong enough to trigger the
subcritical baroclinic instability.  Saturation leads to a fast burst of enstrophy in the
box, and a large-scale vortex develops in the course of the next $\approx$100
orbits. The amount of angular momentum transport achieved is of the
order of $\alpha\approx\ttimes{-3}$, as in compressible SBI models. 
For the first time, a self-sustained 3D vortex is produced from linear 
amplitude perturbation of a quiescent base state. 
\end{abstract}


\section{Introduction}
\label{sect:introduction}

Accretion in disks is generally thought to occur by the action of 
turbulence, for which the magnetorotational instability
\citep[MRI,][]{BalbusHawley91} is the most likely culprit. However, protoplanetary disks are cold; the 
ionization level required to couple the gas to the ambient field is not
always met \citep{BlaesBalbus94}, leading to zones that are ``dead''
to the MRI \citep{Gammie96,TurnerDrake09}. So, the
quest for hydrodynamical sources of turbulence continues, if only to
provide accretion through this dead zone. 

One such possible sources of hydrodynamical turbulence is the
subcritical baroclinic instability 
\citep[SBI,][]{KlahrBodenheimer03,Klahr04,Petersen07a,Petersen07b,LesurPapaloizou10,LyraKlahr11,Raettig13}, 
a process shown to sustain large-scale vortices in the presence of a radial entropy gradient and
thermal relaxation or diffusion. Two-dimensional linear stability
analysis and numerical simulations do not find instability if only
seeded with linear noise \citep{JohnsonGammie05}, though it was 
shown that finite amplitude perturbations would trigger it, 
concluding that the instability is nonlinear in nature \citep{LesurPapaloizou10}. Characterization of the instability through
nonlinear numerical simulations shows that maximum amplification is
found for thermal times in the range of 1--10 times the dynamical timescale \citep{LesurPapaloizou10,LyraKlahr11,Raettig13}. Although no criterion for a
critical Reynolds number was derived, \citet{Raettig13} show that
as resolution is increased, ever smaller perturbations are necessary, as
expected if the process is physical. Compressible simulations
\citep{LesurPapaloizou10,LyraKlahr11} 
show that the spiral density
waves excited by the vortices \citep{HeinemannPapaloizou09a,HeinemannPapaloizou09b,HeinemannPapaloizou12}
transport angular momentum at the level of $\alpha\approx\ttimes{-3}$,
where $\alpha$ is the Shakura-Sunyaev parameter
\citet{ShakuraSunyaev73}. If this process indeed occur in disks, it would provide not only
accretion but also a fast route for planet formation in the dead zone, since vortices
speed up the process enormously, by concentrating particles in their centers
\citep{BargeSommeria95,KlahrBodenheimer06,Lyra08b,Lyra09}.  

The appeal of the SBI, however, is severely hindered by its nonlinear
nature. Without the guide of analytics, nonlinear processes are
difficult to characterize, and the accuracy of the numerics have to
be well-established beyond reasonable doubts. 
Recently, Klahr \& Hubbard (2014, hereafter KH14) have claimed that, 
when considering the same equations that lead to SBI in 2D, linear
growth exists if vertical wavelengths are considered. The unstable mode is a slowly growing
epicyclic oscillation, which led the authors to name the process
``convective overstability''. Growth is powered by buoyancy and thermal
relaxation in the same regime as the SBI, of cooling time of the order of
the dynamical time. We analyze this claim of linearity in more detail in this paper. Independent
verification is desirable since unorthodox assumptions were made in the
linear analysis of KH14. In particular, the authors assumed that the
timescale for pressure equilibration is fast, and thus set the
pressure perturbation to zero in the linear analysis. Because of this
strong assumption, skepticism about the validity of the work naturally remains until a
rigorous derivation of the dispersion relation is provided,
unambiguously demonstrating that the eigenvector of the 
growing root has no appreciable pressure term. In this work, we
provide such derivation. 

Another point raised by KH14 is the connection
between this overstability and the SBI, if any. A priori, the two processes have little
to do with each other. However, as the regimes of cooling time for both
are similar, if the convective overstability exists, it
may generate the finite amplitude perturbations that trigger the
SBI. In this scenario, the (nonlinear) SBI would simply be the saturated state of
the (linear) convective overstability. Since the difficulty on finding
a source of finite amplitude perturbation in dead zones in the
required range of cooling times had made the SBI look less attractive as a 
relevant disk process, a linear process that can spawn the SBI from
arbitrarily low-level noise would be particularly
interesting. Conversely, there is the possibility, of course, 
that the saturated state of the convective overstability 
may still be of too low amplitude to trigger the SBI. We investigate
these possibilities in the present study. 

This paper is structured as follows. In Sect 2 we perform a linear
analysis calculating the full compressible dispersion relation. In
Sect 3 we take the anelastic limit to derive the instability
criterion, finding the roots, the most unstable mode, and associated
eigenvector. In Sect 4 we perform numerical simulations in the shearing
box to characterize the linear growth phase and nonlinear saturation
in 2D and 3D. We conclude in Sect 5. 

\begin{table}
\caption[]{Symbols used in this work}
\label{table:symbols}
\begin{center}
\begin{tabular}{l l l}\hline
Symbol & Definition & Description \\\hline
$r$ & & cylindrical radial coordinate\\
$\phi$ & & azimuth \\
$r_0$ & & reference radius\\
$x$ & $=r - r_0$& Cartesian radial coordinate\\
$y$ & $=r\phi$ & Cartesian azimuthal coordinate \\
$z$ && vertical coordinate\\
$k_r,k_x$ & & radial wavenumber\\
$m$ & & azimuthal wavenumber\\
$k_z$ && vertical wavenumber\\
$k$ & $=\sqrt{k_r^2+k_z^2}$  & \\
$\mu$ &$=k_z/k$ & \\
$t$&  & time \\
$\rho$ & & density\\
$\v{u}$ & & velocity\\
$T$& & temperature\\
$\gamma$ & & adiabatic index\\ 
$\cp$ & & specific heat at constant pressure \\ 
$\cv$ &$=\cp/\gamma$& specific heat at constant volume \\ 
$p$&$=\cv (\gamma-1) \rho T $ & pressure \\
$\tau$ & & thermal time \\
$\varsigma$ &$=\sfrac{1}{\gamma\tau}$ & \\
$\varOmega$ & & Keplerian angular frequency \\
$q$ & $=-d\ln\varOmega/d\ln r $ & shear parameter \\
$\kappa$ & $=\sqrt{2(2-q)} \ \varOmega$ & epicyclic frequency \\
$\alpha$ & $=d\ln\rho/d\ln r $ & density gradient \\
$\beta$ & $=d\ln T/d\ln r $ & temperature gradient \\
$\ksi$ & $=\alpha+\beta$ & pressure gradient\\
$\omega$ && complex eigenfrequency\\
$\bar\omega$ &$=\omega-m\varOmega$& \\
$s$ & $={\rm Re}(\bar\omega)$ & oscillation frequency\\
$\sigma$ & $={\rm Im}(\bar\omega)$ & growth rate \\
$c$ &$=\left[T \ \cp (\gamma-1)\right]^{1/2}$& sound speed\\
$A$ & $=c \ \partial_r \ln \rho$ &  \\
$B$ & $=\gamma^{-1}c\ \partial_r \ln p$ & \\
$N$ & $=\sqrt{AB -B^2}$ & Brunt-V\"ais\"al\"a frequency  \\
$a$ & $=A/c$ & \\
$b$ & $=B/c$ & \\
$H$ &$= c/\varOmega$ & disk scale height \\
$h$ &$=H/r$ & disk aspect ratio \\
$\varSigma$ & $\propto \rho H$ & surface density \\
$\psi$ & $=d\ln \varSigma/d\ln r $ & surface density gradient\\\hline
\end{tabular}
\end{center}
\end{table}

\section{Linear dispersion relation}

Let us consider the compressible Euler equations with thermal relaxation.  

\begin{eqnarray}
\ptderiv{\rho} + \left(\v{u}\cdot\del\right) \rho &=& -\rho\Div\v{u}, \label{eq:continuity}\\
\ptderiv{\v{u}} + \left(\v{u}\cdot\del\right) \v{u} &=& -\frac{1}{\rho} \grad{p} + \v{g},\label{eq:momentum}\\
\ptderiv{p} + \left(\v{u}\cdot\del\right) p &=& -\gamma p\Div\v{u} -\frac{p}{T}\frac{(T-T_0)}{\tau},\label{eq:energy}
\end{eqnarray} \noindent where $\rho$ is the density, $\v{u}$ is the
velocity, $p$ is the pressure, $\gamma$ is the
adiabatic index, $T$ is the temperature, $T_0$ is a reference
temperature, and $\tau$ is the thermal time. We consider the cylindrical
approximation, meaning that we omit the vertical component of the
stellar gravity, as well as vertical stratification. In this
approximation, the gravity is $\v{g}=-\varOmega^2\v{r}$, with
$\varOmega$ the Keplerian angular frequency and $\v{r}$ the cylindrical
radial coordinate. A list of the mathematical symbols used in this work, together
with their definitions, is provided in Table~\ref{table:symbols}. 

We linearize \eqss{eq:continuity}{eq:energy} into base state and
perturbation (the latter denoted by primes), as $u_r = u_r^\prime$, $u_\phi=u_\phi^\prime + \varOmega r$,
$u_z=u_z^\prime$, $p=p_0 + p^\prime$, and $\rho=\rho_0+\rho^\prime$. 
Assuming the cylindrical approximation ($\partial_z$ =0 for the base
state), \eqss{eq:continuity}{eq:energy} become 

\begin{eqnarray}
\partial_{\hat t} \rho^\prime + u_r^\prime\partial_r \rho_0 +\rho_0\Div{\v{u}^\prime} &=&0,\label{eq:cont2}\\
\partial_{\hat t} u_r^\prime -2\varOmega u_\phi^\prime +  \frac{1}{\rho_0}\partial_r p^\prime - \frac{\rho^\prime}{\rho_0^2} \partial_r p_0 &=&0,\\
\partial_{\hat t} u_\phi^\prime +\varOmega (2-q) u_r^\prime + \frac{1}{\rho_0} \partial_{\hat\phi} p^\prime &=&0,\\
\partial_{\hat t} u_z^\prime + \frac{1}{\rho_0}\partial_z p^\prime &=&0,\\
\partial_{\hat t} p^\prime +u_r^\prime \partial_r p_0 + \gamma p_0 \Div{\v{u}^\prime} 
 + \frac{p^\prime}{\tau} - \frac{p_0 \rho^\prime}{\rho_0\tau} &=&0.\label{eq:energy2}
\end{eqnarray}

\begin{figure}
  \begin{center}
    \resizebox{\columnwidth}{!}{\includegraphics{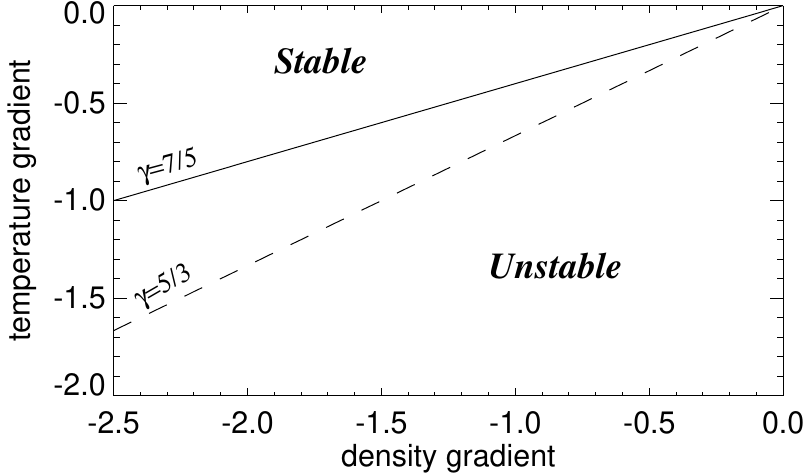}}
\end{center}
\caption[]{The sign of the square of the Brunt-V\"ais\"al\"a frequency
  defines the stability criterion, here shown as a function of the
  density and temperature power-law indices. The plot shows the lines for two
  values of $\gamma$. Above (below) the respective line the system is
  stable (unstable).  } 
\label{fig:stability}
\end{figure} 

\begin{figure*}
  \begin{center}
    \resizebox{\textwidth}{!}{\includegraphics{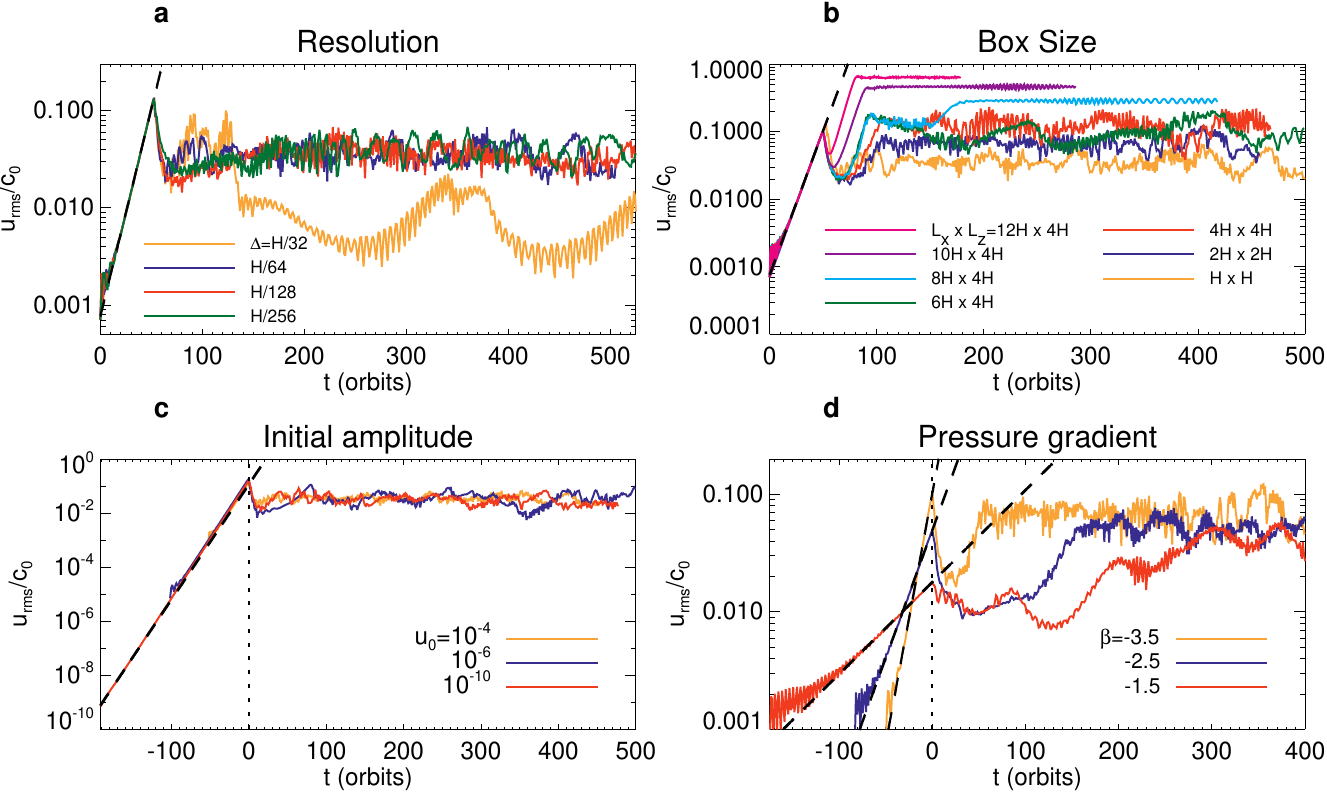}}
\end{center}
\caption[]{Convergence study of the saturated state of the overstability. {\bf a}, the box
size ($H{\times}H$), initial amplitude ($u_{\rm rms}/c=\ttimes{-3}$) and
pressure gradient ($\ksi=-3.5$) were kept fixed, while the resolution
was changed. Saturation occurs at 64 points per scale height. {\bf b}, 
The resolution is fixed at $\Delta=H/64$, and the box size
changed. There is no convergence with box size (see fig 3 and
discussion in the text). {\bf c}, resolution $\Delta=H/64$, box size 
$H{\times}H$, and varying initial amplitude. {\bf d.}, varying the
pressure gradient, resolution $\Delta=H/64$, box size 2H. 
Amplitude converges in both latter cases. The
linear growth rate (black dashed line) is very well reproduced in all cases.}
\label{fig:2D-saturation}
\end{figure*}

\begin{figure*}
  \begin{center}
    \resizebox{\textwidth}{!}{\includegraphics{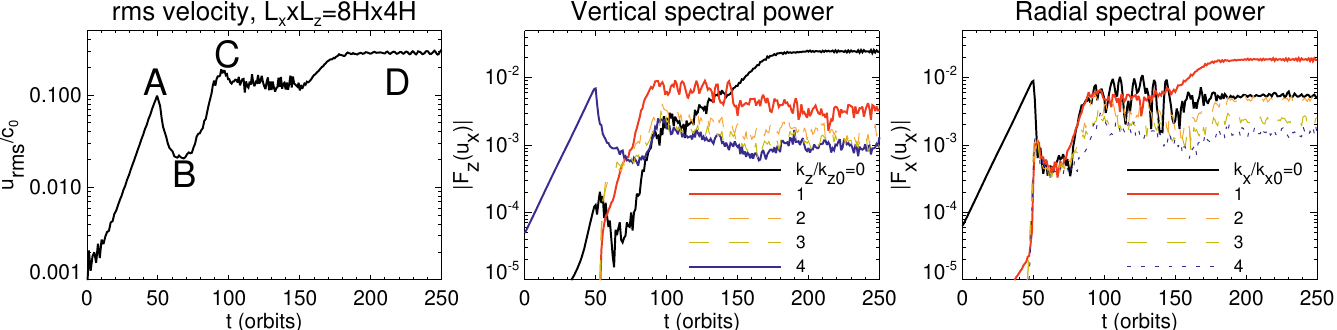}}
    \resizebox{\textwidth}{!}{\includegraphics{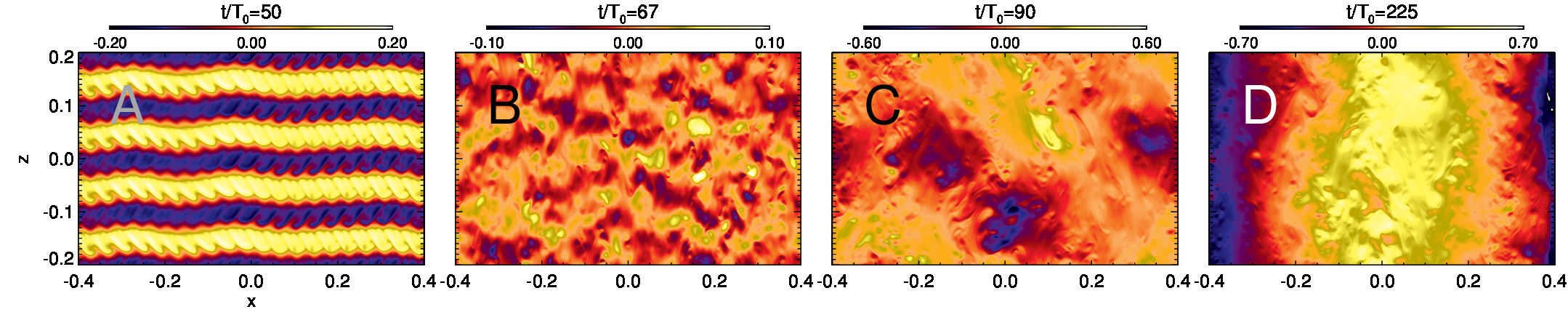}}
\end{center}
\caption[]{Spectral description of the $8H{\times}4H$ simulation (cyan line in fig
  2, upper right). The rms velocity is shown in the upper left panel,
  with four representative points marked, and the velocity shown, in
  the xz plane, in each of these points (lower panels). The points
  are: $A$, onset of saturation; $B$, the local
  minimum; $C$, second saturated state; $D$, state after the last
  bifurcation. The spectral power in the first 5 large-scale modes in
  $x$ and $z$ is shown in the upper middle and right panels,
  respectively. Point $A$ corresponds to Kelvin-Helmholtz instability breaking
  up the original $k_z/k_{z0}=4$ channel mode, as nonzero $k_x$ and
  $k_z$ modes are excited. The local minimum $B$ corresponds to the
  point when power is equally distributed among the non-zero $k_z$
  modes. The saturated state $C$ corresponds to dominance of the $k_z=1$
  mode, with a mixed $k_x=1$ mode (both clear in the lower ``C''
  panel). The last bifurcation corresponds to the $k_z=0$ mode 
  taking over, and a large scale $k_x=1$ dominating the box. }
\label{fig:power}
\end{figure*}

In the above equations, $\partial_{\hat t} = \partial_t +
\varOmega\partial_\phi$, $\partial_{\hat\phi}=r^{-1} \ \partial_\phi$,
$\Div{\v{u}^\prime}=\partial_r u_r^\prime + u_r^\prime/r
+ \partial_{\hat\phi} u_\phi^\prime + \partial_z u_z^\prime$, and the thermal
relaxation term was linearized \[ \frac{\delta T}{T} = \frac{\delta p}{p} -
\frac{\delta \rho}{\rho},\] as per the equation of state, $p=c_v(\gamma-1) \rho
T$. Next we use the short-wave approximation, $m \ll k_rr, k_z z$, and expand the
perturbations in Fourier modes, $\exp(-i\omega t + ik_rr + im\phi +
ik_z z)$. \eqss{eq:cont2}{eq:energy2} then become

\begin{eqnarray}
-i\bar\omega \rho^\prime + u_r^\prime \partial_r \rho_0 + \rho_0 i k_ru_r^\prime + \rho_0ik_z u_z^\prime&=& 0 \\
-i\bar\omega u_r^\prime - 2\varOmega u_\phi^\prime + ik_r\rho_0^{-1} p^\prime -\frac{\rho^\prime}{\rho_0^2}\partial_r p_0 &=& 0 \\
-i\bar\omega u_\phi^\prime + \varOmega(2-q) u_r^\prime &=& 0 \\
-i\bar\omega u_z^\prime + ik_z\rho_0^{-1} p^\prime &=&0\\
-i\bar\omega p^\prime + u_r^\prime \partial_r p_0 + \rho_0 c^2ik_ru_r^\prime + \rho_0c^2 i k_z u_z^\prime &&\nonumber\\
+ \frac{p^\prime}{\tau} -\frac{c^2 \rho^\prime}{\gamma\tau}&=& 0 
\end{eqnarray}

\noindent where $\bar\omega=\omega-m\varOmega$, and we have also
substituted $p_0 = \rho_0 c^2 / \gamma$. The system is
$\vt{M}\cdot\v{\upsilon}=0$, where
$\v{\upsilon}=[\rho^\prime,u_r^\prime,u_\phi^\prime,u_z^\prime,p^\prime]^T$,
and the coefficient matrix is 

\begin{equation}
\vt{M}=\left[\begin{array}{ccccc}
      -i\bar\omega&\rho_0 (i k_r+\sfrac{A}{c}) & 0 & \rho_0 i k_z & 0 \\
      -\sfrac{B c}{\rho_0}&-i\bar\omega&-2\varOmega&0 & \sfrac{ik_r}{\rho_0}\\
      0&\varOmega(2-q)&-i\bar\omega&0&0\\
      0&0&0&-i\bar\omega & \sfrac{ik_z}{\rho_0}\\
      -\sfrac{c^2}{\gamma\tau}&\rho_0 c^2 (i k_r + \sfrac{B}{c})& 0 &\rho_0 c^2 i k_z & -i\bar\omega + \sfrac{1}{\tau}\\
    \end{array}\right].\label{eq:matrix}
\end{equation}

\noindent We have substituted  

\begin{eqnarray}
A&=&\white{\gamma^{-1}}c \ \partial_r\ln \rho,\\
B&=&\gamma^{-1} c \ \partial_r\ln p,
\end{eqnarray}
\noindent so both $A$ and $B$ have dimension of frequency. In particular,
$AB=1/\rho^2 \partial_r\rho \partial_rp$, 
and $B^2 = (\gamma \rho p)^{-1} (\partial_rp)^2$, so $N^2=AB-B^2$ is the
square of the Brunt-V\"ais\"al\"a frequency. The full dispersion
relation ${\rm det}\vt{M}=0$ is

\begin{eqnarray}
&&\bar{\omega}^5 + \bar{\omega}^4 \ i\tau^{-1} - \bar{\omega}^3  \ (AB +  c^2k^2 + \kappa^2) \nonumber\\
&+&\bar{\omega}^2 \ \tau^{-1}\left[ k_rc (B-A\gamma^{-1})  - i( AB + c^2k^2\gamma^{-1} + \kappa^2)\right] \nonumber\\
&+& \bar{\omega} \ c^2k_z^2(\kappa^2+N^2) +
\frac{ic^2\kappa^2k_z^2}{\gamma\tau} = 0, 
\label{eq:dispersion}
\end{eqnarray}

\noindent where $\kappa^2 = 2(2-q)\varOmega^2$ is the square of the
epicyclic frequency. We consider now some limits of \eq{eq:dispersion}.

\section{Anelastic limit}

In the anelastic limit, $c=\infty$, \eq{eq:dispersion} reduces to 

\begin{eqnarray}
\bar{\omega}^3  k^2 - \bar{\omega}^2 \ \tau^{-1}\left[k_r(b-a/\gamma)  - ik^2\gamma^{-1}\right] &&\nonumber\\
- \bar{\omega} \ k_z^2(\kappa^2+N^2) - \frac{i\kappa^2k_z^2}{\gamma\tau} &=& 0,
\label{eq:anelastic-1}
\end{eqnarray}

\noindent where $b=B/c =\gamma^{-1} \partial_r \ln p$ and
$a=A/c=\partial_r\ln\rho$. These terms are proportional to $1/r$,
so they are small and can be dropped. The dispersion relation is thus

\begin{equation}
\bar{\omega}^3  + i\varsigma\bar{\omega}^2 - \bar{\omega} \mu^2(\kappa^2+N^2) - i\varsigma\kappa^2\mu^2 = 0, 
\label{eq:anelastic}
\end{equation}

\noindent where we have also substituted $\varsigma=1/\gamma\tau$ and $\mu^2=k_z^2/k^2$.

\subsection{Adiabatic}

For adiabatic flow, $\tau=\infty$, \eq{eq:anelastic} reduces to 

\begin{equation}
  \bar\omega^2 = \mu^2 \left( \kappa^2+ N^2\right)
\end{equation}

\noindent For $k_r=0$ (in-plane incompressible motion), we retrieve
$\bar{\omega}^2=\kappa^2+N^2$, the Solberg-Hoiland criterion.

\subsection{Finite $\tau$, $k_r=0$}

For pure in-plane incompressible motions ($k_r=0$), \eq{eq:anelastic} reduces to 

\begin{equation}
  \bar\omega^3  + \bar\omega^2i\varsigma -\bar\omega (\kappa^2+ N^2)
  - i\varsigma \kappa^2 = 0 ,
\end{equation}

\noindent which is the same as derived by KH14 (their eq. 18), using other assumptions.

\subsection{Finite $\tau$, $k_r \ne 0 $}

Substituting $\bar{\omega}=s + i\sigma$, growing solutions
correspond to real positive $\sigma$. The dispersion relation, real and imaginary, that  need to vanish
independently, are:  

\begin{equation}
s^2 = \mu^2(N^2+\kappa^2) + 3\sigma^2  + 2\sigma \varsigma;
\label{eq:s}
\end{equation}

\begin{equation}
  \sigma^3  +\sigma^2 \varsigma - \sigma[3s^2 - \mu^2(N^2+\kappa^2)] - \varsigma(s^2 -\mu^2\kappa^2) = 0. 
  \label{eq:gamma}
\end{equation}

Substituting \eq{eq:s} into \eq{eq:gamma}, we get 

\begin{equation}
2\sigma (2\sigma + \varsigma)^2 + 2\sigma \mu^2(\kappa^2+N^2) +
\mu^2\varsigma N^2 = 0. 
\label{eq:full-dispersion}
\end{equation}

As we expect the growth to be small (to be checked a posteriori), we take the limit $\sigma
\ll \varsigma$, leading to 

\begin{equation}
\sigma =-\frac{1}{2}\left[\frac{\mu^2\varsigma N^2}{\varsigma^2 + \mu^2(\kappa^2+N^2)} \right].
\end{equation}

This function has no extrema for finite $\mu$. For $\varsigma$, however, there
is a maximum at $\varsigma^2\vert_{_{d_t\sigma=0}} = \varsigma_{\rm max}^2=\mu^2(\kappa^2+N^2)$, that is, maximum growth
occurs for 

\begin{equation}
  \tau_{\rm max} = \frac{1}{\gamma}\left|\frac{k}{k_z}\right|\frac{1}{\sqrt{\kappa^2+N^2}}
\end{equation}

\noindent for which the growth rate is $\sigma_{\max} = -\mu^2N^2/(4
\varsigma_{\rm max})$, i.e.  

\begin{equation}
  \sigma_{\rm max} = -\frac{1}{4}\left|\frac{k_z}{k}\right|\frac{N^2}{\sqrt{\kappa^2+N^2}}
\end{equation}

\subsubsection{Keplerian disks}

Recalling the definition of the Brunt-V\"ais\"al\"a frequency 

\begin{equation}
  N^2 \equiv \frac{1}{\rho}\frac{dp}{dr}\left(\frac{1}{\rho}\frac{d\rho}{dr} - \frac{1}{\gamma p}\frac{dp}{dr}\right),\end{equation}

\noindent we can write it in terms of the power-law indices of the density and temperature
gradients, $\alpha=\partial\ln\rho/\partial\ln r$, 
$\beta=\partial\ln T/\partial\ln r$, and $\ksi=\alpha+\beta = \partial\ln p/\partial\ln r$, resulting in  

\begin{equation}
N^2 = \frac{\varOmega^2 h^2}{\gamma}\left( \alpha\ksi - \frac{1}{\gamma}\ksi^2\right),
\end{equation}

\noindent where $h=H/r$ is the aspect ratio and $H=c/\varOmega$ is the
scale height. So, for Keplerian disks, 
$\kappa=\varOmega$ and  $|N^2| \sim \varOmega^2
\mathcal{O}(h^2)$. It results from this that $\tau_{\rm max}$ is of order $1/\varOmega$, while the
associated growth rate is of order $\sigma_{\rm max} = \varOmega \mathcal{O} (h^2)$,
validating the assumption that $\sigma \ll\varsigma$. 

Notice that for $k_r\gg k_z$, that is, $\mu^2\rightarrow 0$, the dispersion relation \eqp{eq:full-dispersion} becomes 

\begin{equation}
  \sigma(2\sigma+\varsigma)^2 = 0, 
\end{equation}

\noindent for which the roots are $\sigma=0$, and $\sigma=-\varsigma/2$, that is, no
growth, and damped perturbations. For channel modes ($k_r=0$) in
Keplerian disks ($\kappa=\varOmega\gg|N|$), we find 

\begin{eqnarray}
  \tau_{\rm max} &=& \white{+}\frac{1}{\gamma\varOmega};\\
  \sigma_{\rm max} &=& -\frac{N^2}{4\varOmega}.
\end{eqnarray}

We plot in \fig{fig:stability} the unstable range as a function
of the density and temperature power law indices. {\footnote{Notice that the
condition that $N^2 < 0$ requires (for $\ksi<0$) that $\alpha -
\sfrac{\beta}{(\gamma-1)} > 0$. For a power-law surface density
$\varSigma\propto \rho H \propto r^\psi$, we have $\psi =
\alpha+\sfrac{\beta}{2}+\sfrac{3}{2}$. The requirement is then 
$2\psi > 3 + \beta \ \sfrac{(\gamma+1)}{(\gamma-1)}$, 
which, for $\gamma=\sfrac{7}{5}$ means $\psi > 3(\beta+\sfrac{1}{2})$. 
For $\beta=-\sfrac{1}{2}$ the surface density has to be flat or
increasing with distance in order to lead to instability, which is
not reasonable. For $\beta=-\sfrac{3}{4}$ the onset of instability corresponds
to $\psi=-\sfrac{3}{4}$ (also for $\gamma=\sfrac{7}{5}$), 
which is consistent with the range of $\psi\approx [-0.4,-1.0]$ (with median -0.9) 
found in the observations of \citet{Andrews09}.}}

\subsection{The unstable mode}

To understand the most unstable mode, we check the eigenvector
$\v{\upsilon}_{\rm max}$
corresponding to this root, for which the eigenvalue is 

\begin{equation}
\lambda =  i\omega_{\rm max} = i\varOmega -\sigma_{\rm max},
\end{equation}

\noindent and the system is $\vt{R}\cdot\v{\upsilon}_{\rm max} =
\lambda\v{\upsilon}_{\rm max}$, where

\begin{equation}
\vt{R}=\left[\begin{array}{ccccc}
      0&\rho_0 \sfrac{A}{c} & 0 & \rho_0 i k_z & 0 \\
      -\sfrac{B c}{\rho_0}&0&-2\varOmega&0 & 0\\
      0&\varOmega/2&0&0&0\\
      0&0&0&0 & \sfrac{ik_z}{\rho_0}\\
      -c^2\varOmega&\rho_0 Bc& 0 &\rho_0 c^2 i k_z & \gamma\varOmega\\
    \end{array}\right].\label{eq:matrix}
\end{equation}

The 4th line is $ik_z/\rho_0 u_z^\prime = i\varOmega
u_z^\prime -\sigma_{\rm max} u_z^\prime$, which is only satisfied for
the trivial solution $u_z^\prime$=0. The reduced system becomes 

\begin{eqnarray}
\rho^\prime &=& \lambda^{-1} \rho_0 a \ u_r^\prime;\\
u_\phi^\prime &=& \lambda^{-1}\varOmega/2 \ u_r^\prime;\\
p^\prime &=& (\lambda - \gamma\varOmega)^{-1} \left(\rho_0 c B u_r^\prime -c^2\varOmega \rho^\prime \right).
\end{eqnarray}

The solution is 

\begin{eqnarray}
\ln\rho^\prime &=& -\frac{\sigma+i\varOmega}{\varOmega} a u_r^\prime; \\
u_\phi^\prime &=& -\frac{\sigma+i\varOmega}{2\varOmega} u_r^\prime;\\
\ln p^\prime &=& -\gamma\left(\frac{\sigma}{\varOmega}a + b\right)\left[ \frac{\sigma +\varOmega\gamma + i\varOmega}{(\sigma+\gamma\varOmega)^2+\varOmega^2} \right] u_r^\prime.
\end{eqnarray}

Since $\sigma\ll\varOmega$, the pressure perturbation is 

\begin{equation}
\ln p^\prime = -\frac{\gamma(\gamma+i)}{\varOmega(\gamma+1)} b  u_r^\prime.
\end{equation}

And, because $a$ and $b$ are or order $\sfrac{1}{r}$, $\ln \rho^\prime$ and $\ln p^\prime$ are
vanishingly small. That the pressure variation does
not play a major role in the instability justifies (now a posteriori) the $p^\prime=0$
approximation of KH14. The eigenvector is simply 
\begin{equation}
  \v{\upsilon}_{\rm max}=\left[0,1,-\frac{1}{2}\left(\frac{\sigma}{\varOmega}+i\right),0,0\right]^T,
  \label{eq:eigenvector}
\end{equation}

\noindent i.e., an overstable epicycle.

\begin{figure*}
  \begin{center}
    \resizebox{.98\textwidth}{!}{\includegraphics{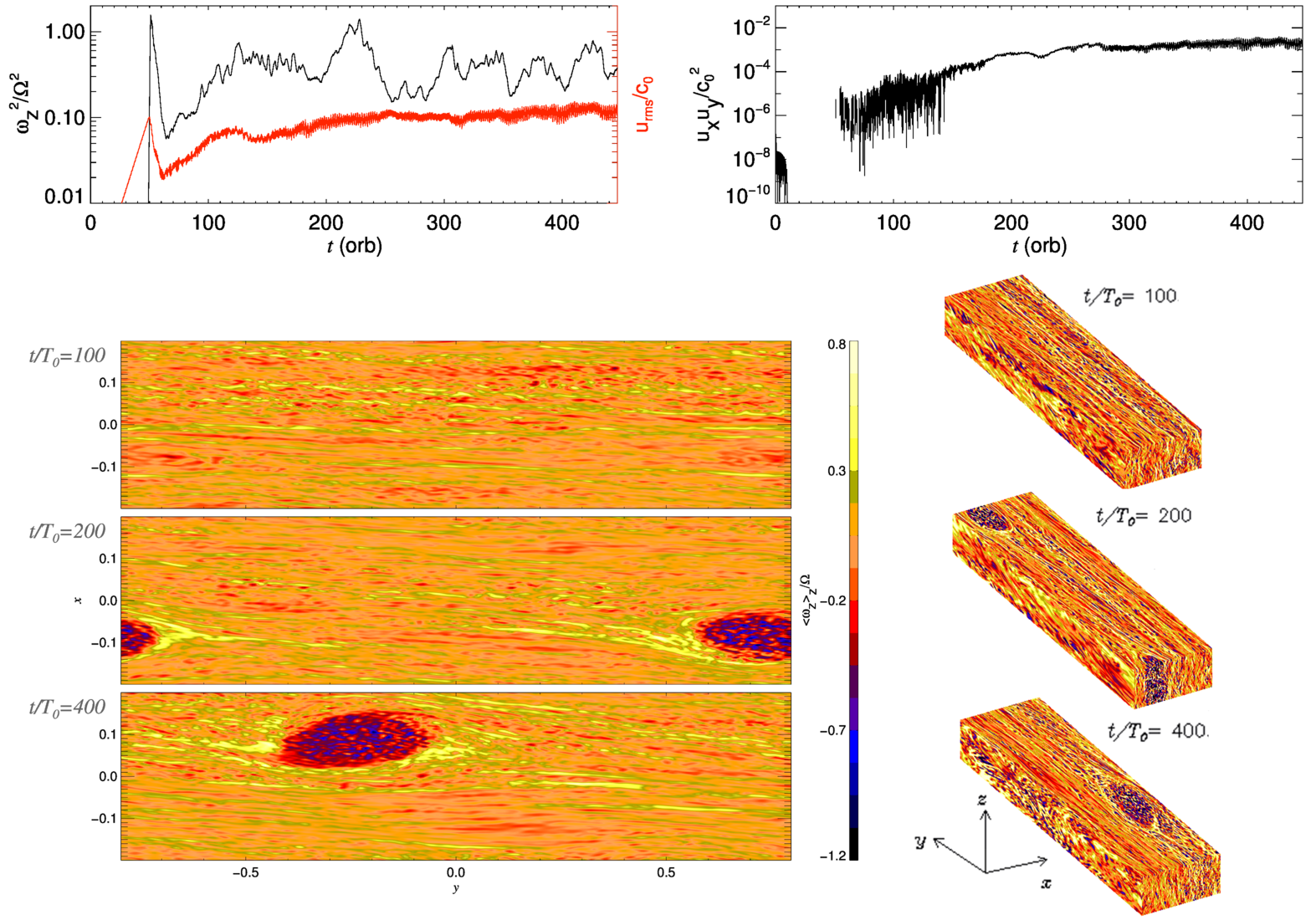}}
\end{center}
\caption[]{Nonlinear evolution of the buoyant overstability in three
  dimensions. With the linear overstability raising the amplitude of
  the initial fluctuations to nonlinear levels, the saturated state is
  expected to be similar to that of the SBI. In the lower left panels
  we show the averaged vertical vorticity; the lower right shows the
  vorticity in the 3D flow. Indeed, we see that large-scale
  self-sustained anticyclonic vortices develop in
  the saturated state. The upper left panel shows the radial velocity rms
  and vertical enstrophy (red and black line, respectively.) The upper
  right panel shows the level of Reynolds stress, saturating at $\alpha\approx\ttimes{-3}$.} 
\label{fig:figure3d}
\end{figure*}

\section{Numerical simulations}

We now turn to numerical simulations to check the evolution of the
instability. We use the shearing box model of \citet{LyraKlahr11},
that includes the linearized pressure gradient. We do so in 
order to benefit from shear-periodic boundaries, in contrast to the simulations 
in the appendix of KH14, that are affected by radial boundaries. 
The reader is referred to \citet{LyraKlahr11} for the 
equations of motion, properties and caveats of 
the approximation. In particular, the density gradient is zero, and we
drop the $x$-dependent term in the pdV work to keep shear-periodicity
(see appendix A of \citet{LyraKlahr11}. 

We solve the evolution equations with 
the {\sc Pencil Code} \citep{BrandenburgDobler02} {\footnote{The code,
including improvements done for the present work, is publicly
available under a GNU open source license and can be downloaded at
http://www.nordita.org/software/pencil-code}} which integrates
the PDEs with sixth order spatial derivatives, and
a third order Runge-Kutta time integrator. Sixth-order
hyper-dissipation terms are added to the evolution equations,
to provide extra dissipation near the grid scale, explained in
\citet{Lyra08a}. They are needed because the high-order scheme of
the Pencil Code has little overall numerical dissipation 
\citep{McNally12}.

We run a suite of 2D axisymmetric models ($x$ and $z$) to understand
the linear evolution and saturation properties of the instability. 
The sound speed is $c$=0.1, and the adiabatic index is $\gamma=1.4$. The cooling time is 
$\tau_{\rm  max}=\sfrac{1}{\gamma\varOmega}$. Our units are
$\varOmega=c_p=\rho_0=1$. 

We initialize the simulations with the eigenvector corresponding to
the epicycle oscillation, $u_y=-iu_x/2$. Because for $k_x=0$ the growth rate does
not depend on $k_z$, we arbitrarily choose $\lambda_z=H$ for the
channel mode. The initial condition therefore is 

\begin{eqnarray}
u_x&=&u_0\sin\left(\frac{2\pi}{H} z\right); \\
u_y&=&\frac{u_0}{2} \cos\left(\frac{2\pi}{H} z\right).
\end{eqnarray}

The fiducial model has resolution $\Delta{x}=\Delta{z}=H/64$, box size
$L_x\times L_z=2H\times2H$,
temperature gradient $\beta=-3.5$, and initial amplitude $u_0/c
=\ttimes{-3}$. We vary these quantities to check convergence at
saturation. The evolution of the 2D axisymmetric box seeded with the
channel mode is shown in the panels of \fig{fig:2D-saturation}. The linear phase
matches the analytical prediction (dashed black line) for all
models ran. 

Figure~\ref{fig:2D-saturation}a shows the dependency on
resolution. Convergence is achieved for 64 grid points per scale
height. There is also convergence for initial amplitude of
perturbation, as seen in \fig{fig:2D-saturation}c. The linear phase is
identical in the three cases examined ($u_0=\ttimes{-4},\ttimes{-6}$,
and $\ttimes{-10}$). In this figure we set $t=0$ as the time that
saturation is achieved, to better compare the nonlinear evolution. In
\fig{fig:2D-saturation}d we check how the instability depends on the
pressure gradient. Again, the linear phase is reproduced for the
different values of the Brunt-V\"ais\"al\"a frequency, and the
amplitudes at saturation are similar, within a factor 2--3.
Difference in seen when we test the dependency on box size
(\fig{fig:2D-saturation}b). The amplitude seemed to saturate
at $4H{\times}4H$ (red line), since the model with $L_x=6H$ (green line) shows a
similar amplitude. However, the model with $L_x=8H$ (cyan line) shows a bifurcation at
$\approx$150 orbits. Models with larger radial range ($L_x=10$ and $L_x=12$,
purple and magenta lines, respectively) show no convergence, even as the
velocity dispersion increasingly approaches the sound
speed. 

Interesting features are seen in this simulation,
that help understand the behavior of the system. We plot in \fig{fig:power}
the time evolution of the power in the first 5 large scale modes, in
both $x$ (upper middle panel) and $z$ (upper left panel). The upper
left panel shows the rms of the radial velocity. Four special/representative instants
are labeled, and the $u_x$ field for these respective instants are
shown in the lower panels. 

The first instant, $A$, corresponds to the first  ``saturation'' seen 
at 50 orbits. The power spectrum shows that 
the clean initial channel mode ($k_z/k_{z0}$=4,  $k_x/k_{x0}$=0)
persisted until this time, after which it saturates, exciting $k_x
\neq 0$ modes and other $k_z$ modes. Instant $B$, at 67 orbits, corresponds to the
local minimum in rms velocity. The power spectrum shows that this
happens when the $k_z/k_{z0}$=1 mode becomes dominant. 
Subsequently, this mode keeps growing, at the same rate as the initial 
$k_z/k_{z0}$=4 mode. This is because the growth rate is independent 
of $k_z$ for $k_x=0$, which at that time has similar power as the
higher $k_x$ modes. From time $t$=90 (instant $C$) to 160 orbits the system settles
into a steady state, with a dominant $k_z/k_{z0}$=1 mode, and mixed
$k_x/k_{x0}$=0 and $k_x/k_{x0}$=1. Another bifurcation happens when
the $k_z=0$ mode overtakes the $k_z/k_{z0}$=1 mode. Simultaneously, 
it prompts $k_x/k_{x0}$=1 to dominate over $k_x=0$. The final state
(labeled $D$) is thus vertically symmetric, with a box-wide radial wavelength. 

This explains why we do not find convergence while increasing box 
vertical range from $L_z$=$H$ to $2H$ to $4H$. In these boxes, because
we kept the seed mode at $k_z=2\pi/H$, we initialized the instability with
the $k_z/k_{z0}$=1, 2, and 4 mode, respectively. In the last two
simulations, the $k_z/k_{z0}$=1 mode was growing, with less power, but
eventually catching up as the seed mode saturates. Convergence with radial box size is never achieved in the 2D
runs because the $k_x/k_{x0}=1$ mode comes to
dominate, no matter how wide we make the box. The simulations
with radial box size $L_x=10H$ and $L_x=12H$ show the same pattern,
albeit with no intermediate phase of dominance of a $k_z/k_{z0}$=1
mode. 

\subsection{3D instability: growth of large-scale vortices}

Next we turn to the 3D evolution of the instability. We set a box of 
size $4H\times 16H\times 2H$, with resolution $256\times256\times128$ in 
$x$, $y$, and $z$, respectively. The cells thus have aspect ratio
$1\times 4\times 1$ (we have checked in \citealt{LyraKlahr11} that
unit aspect ratio in $x$ and $y$ gave the same results for the
twodimensional SBI). 

With the azimuthal direction present, vertical vorticity 
(in-plane circulation) can evolve unabridged. We show in 
\fig{fig:figure3d} (left panel) the evolution of the rms velocity (red line) and
enstrophy (black line). When the initial $k_z$ mode saturates (at 50
orbits, as in the 2D meridional models of fig 3), a sharp rise in enstrophy
occurs. The situation is now very similar to the SBI, with high-amplitude
perturbations ($u_{\rm rms}\approx0.1 c_s$), thermal relaxation, and
an entropy gradient. The nonlinear saturation state of this buoyant
overstability should thus proceed very similarly to the evolution of the 
 SBI. Indeed, as the lower panels of \fig{fig:figure3d} show, the saturated state develops into a large scale
 vortex. The amount of angular momentum transport (\fig{fig:figure3d},
 upper right) is at the $\alpha\approx \ttimes{-3}$ level, again, the typical level of the
SBI. It seems conclusive that the saturated state of the buoyant overstability is the SBI. 

\section{Conclusions} 

We conclude that indeed there is a linear overstability in the region
of the parameter space of negative $N^2$, finite cooling time $\tau$,
and non-zero $k_z$ perturbation. The approximation $\delta p =0$ done
by KH14 is justified as $\delta p =0$ (and $\delta\rho$) in the
eigenvector of the most unstable modes is vanishingly small in comparison 
to the velocity amplitude \eqp{eq:eigenvector}. 

Modeling the system numerically, we reproduce the linear growth rate
in all cases. In the twodimensional meridional simulations, we find convergence in the saturated
state with resolution, but not with box size, since a large-scale 
$k_x/k_{x0}=1$ radial mode dominates the box. However, in three dimensions
this mode does not show up, as it gets sheared away. 

We also show that the SBI is indeed the saturated state of the
overstability. Saturation leads to a fast burst of enstrophy in the
box, and a large-scale vortex develops in the course of the next $\approx$100
orbits after the convective overstability has built the finite
amplitude perturbations. The amount of angular momentum 
transport achieved is of the order of $\alpha\approx\ttimes{-3}$, as in compressible SBI models. 

It remains to be shown if these processes (both SBI and convective
overstability) operate in global models, i.e., how they respond to boundary
conditions and curvature terms. The relation between this
overstability and the Goldreich-Schubert-Fricke instability
\citep{GoldreichSchubert67,Fricke68,Nelson13} 
should also be the subject of future work.

\acknowledgments This work was performed in part at the Jet Propulsion
Laboratory, under contract with the California Institute of Technology
funded by the National Aeronautics and Space Administration 
(NASA) through the Sagan Fellowship Program executed by the NASA 
Exoplanet Science Institute. This paper started from a 
\href{https://www.facebook.com/groups/149092965243653/302839956535619/?notif_t=group_comment#}{discussion} between the author, Alexander Hubbard, Matthew Kunz, Hubert Klahr, Henrik Latter, Geoffroy Lesur, Min-Kai Lin, George Mamatsashvili, and Orkan Umurhan. It further profited from input from Anders Johansen, Mordecai-Mark Mac Low, Colin McNally, Neal Turner, and Andrew Youdin.

\end{document}